\newcommand{\ignore}[1]{}
\documentstyle[epsf,11pt]{article}
\def\be{\begin{equation}}
\def\ee{\end{equation}}
\def\bea{\begin{eqnarray}}
\def\eea{\end{eqnarray}}

\def\be{\begin{equation}}
\def\ee{\end{equation}}
\def\bea{\begin{eqnarray}}
\def\eea{\end{eqnarray}}
\newcommand\half{{\textstyle{1\over2}}}

\begin{document}
\begin{titlepage}
\noindent BROWN-HET-1118:TA556  \hfill April 21, 1998 \\
\begin{center}

{\Large\bf Diffraction Association and Flavoring of Pomeron}

\vspace{3cm}
{\bf Chung-I Tan$^{(1)}$}\\
\end{center}
\vspace{1.0cm}
\begin{flushleft}
{}~~$^{(1)}$Department of Physics, Brown University, Providence, RI 02912,
USA\\
\end{flushleft}
\vspace{3cm}


\abstract{The most important consequence of Pomeron
being a pole is the factorization property.
 However, due to 
Pomeron intercept being greater than 1,  the extrapolated 
single diffraction dissociation cross section based on a classical triple-Pomeron formula is too
large leading to a potential unitarity violation at Tevatron energies.  
 It is
our desire here to point out that  the ``flavoring" of  Pomeron plays the
dominant role in resolving this apparent ``paradox".   }

\vfill
\noindent Talk presented at LAFEX Workshop on Diffractive
Scattering (LISHEP98), Rio de Janeiro, Brasil, Feb. (1998).
\end{titlepage}

\section{Introduction}
One of the more interesting developments  from
recent collider experiments 
 is the finding that  hadronic  total cross sections as well as  elastic cross sections in the
near-forward limit can be  described by the exchange of a ``soft Pomeron"
pole,~\cite{Tan1} {\it i.e.},
 the absorptive part of the elastic amplitudes can be approximated by 
${Im}\> T_{a,b}(s,t)\simeq \beta_a(t) s^{\alpha_{\cal P}(t)}\beta_b(t).$
 The Pomeron trajectory has two important features. First, its
zero-energy intercept is greater than one,
$\alpha_{\cal P}(0)\equiv 1+
\epsilon$,
$\epsilon\simeq 0.08\sim 0.12$, leading to rising $\sigma^{tot}(s)$. Second, its Regge
slope is approximately $\alpha_{\cal P}'\simeq 0.25\sim 0.3$ $
GeV^{-2}$, leading to  
the observed  shrinkage effect for  elastic peaks. 
The most important consequence of Pomeron
being a pole is  factorization. For a singly diffractive dissociation process,
factorization leads to a  ``classical triple-Pomeron" formula,~\cite{Classical} 
$
{d\sigma \over dtd\xi}\rightarrow {d\sigma^{classical} \over dtd\xi}\equiv  F_{{\cal P}/a}^{cl}
(\xi, t)
\sigma_{{\cal P}b}^{cl} (M^2,t),
$
where $M^2$ is the missing mass variable and $\xi\equiv M^2/s$.   The first term, $ F_{{\cal P}/a}^{cl}
(\xi, t)$, is the so-called ``Pomeron flux", and the second term is the
``Pomeron-particle" total cross section.
With  $\epsilon\sim 0.1$, it has been observed~\cite{Dino1} that  the extrapolated
$p\bar p$ single diffraction dissociation cross section, $\sigma^{sd}$,  based on
the standard triple-Pomeron formula is too large at Tevatron energies by as much as a  factor of
$5\sim 10$ and it could   become larger than the
total cross section. 

Let us denote  the singly diffractive cross section  as a product
of a ``renormalization" factor and the classical formula,
\be
{d\sigma \over dtd\xi}=Z(\xi,t;s) {d\sigma^{classical} \over dtd\xi}.
\label{eq:TotalRenormalization}
\ee
Several phenomenological suggestions for this modification factor have been made:
\begin{itemize}
\item{It was  argued by K. Goulianos~\cite{Dino1} that
agreement with data could be achieved by having an energy-dependent suppression factor,
$$Z(\xi,t; s)\rightarrow  N(s)^{-1}\leq 1,$$
so that a new ``Pomeron flux", 
{$F_N(s, \xi, t) \equiv N(s)^{-1}F_{{\cal P}/p}^{cl} (\xi, t)$}, is always normalized  
 to be less than unity: $N(s)\simeq  (s/\bar s)^{2\epsilon}>1$  for {$ s\geq \bar s$}, and $N(s)=1$ for
 $s< \bar s$, where { $\sqrt{\bar
s}\simeq 22\> GeV$}.}
\item {   
 An alternative suggestion has  been made recently by P.
Schlein,~\cite{Schlein1} by introducing a ``flux damping" factor, 
$$Z(\xi,t; s)\rightarrow D(\xi).$$
This factor has the following features: (i) $D(\xi)\sim 0(1)$, for $1>
\xi>\xi_1$, (ii) $D(\xi)$ drops by a factor of 2 as $\xi$ decreases from $\xi_1$ to $\xi_2$, and (iii) 
$D(\xi)\rightarrow 0$ rapidly for 
$0\leq \xi<\xi_2$, with $\xi_1\sim 0.015$ and $\xi_2\sim 10^{-4}$.}
\end{itemize}

 In view of the factorization property for total and elastic cross sections, the ``flux
renormalization"  procedure, which breaks factorization, appears paradoxical.  On the other hand, the
occurrence of unusaul scales for the Schlein damping factor, e.g., $\xi_2\sim 10^{-4}$, appears
equally mysterious.  We shall refer
 to this
as ``{\bf Dino's paradox}''.
 {Finding a resolution  consistent with Pomeron
pole dominance for elastic and total cross sections at Tevatron energies will be the main focus of
this study.}~\cite{Tan1}

A natural expectation for the
resolution to this paradox  lies in  implementing a large screening correction to the
classical triple-Pomeron formula.  However, this appears too simplistic. In the absence of a new
energy scale, a screening factor of the order $5\sim 10$, if obtained, would apply both at
Tevatron energies and at ISR energies. This indeed is the case for the eikonalization analysis by
Gotsman, Levin, and Maor,~\cite{GotsmanLevinMaor} as pointed out by Goulianos. Since a successful
triple-Pomeron phenomenology exists up to   ISR energies, a subtler explanation is required. We shall
assume that any screening effect can supply at the most a
$10 \sim 20\%$ suppression and it cannot serve as the primary mechanism for explaining the paradox.

Triple-Regge phenomenology has had a long history. It has enjoyed many
 successess since early seventies, and it should emerge as a feature of any realistic
representation of non-perturbative QCD for high energy scattering. In particular, it should be recognized
that, up to ISR energies, triple-Pomeron phenomenology has provided a successful description for the
 phenomenon of diffractive dissociation.
A distingushing feature of the successful low-energy  triple-Pomeron analyses is the value
of the Pomeron intercept. It has traditionally been taken to be near $1$, which would lead to total
cross sections having constant ``asymptotic values". In contrast, the current paradox centers around
the Pomeron having  an intercept great
than 1, e.g.,
$\epsilon\simeq 0.1$.  

Instead of trying to ask ``how can one obtain a large suppression factor at Tevatron energies", an
alternative approach can be adopted. We could first determine the ``triple-Pomeron" coupling by
matching the diffractive cross section at the highest Tevatron energy. A naive
extrapolation to lower energies via a standard triple-Pomeron formula would of course lead to too
small a  cross section at ISR energies. We next ask
the question: 
\begin{itemize}
\item
Are there physics which might have been overlooked by others in moving down in
energies? 
\item {In particular, how can a high energy fit be smoothly interpolated with the successful low
energy triple-Pomeron analysis  using a Pomeron with intercept at 1, i.e.,  $\epsilon\simeq 0$.}
\end{itemize}

A key observation which will help in understanding our proposed resolution concerns the fact that,
even at Tevatron energies, various ``subenergies", e.g., the missing mass squared, $M^2$, and the
diffractive ``gap", $\xi^{-1}$, can remain relatively small, comparable to the situation of ISR
energies for the total cross sections. Our analysis has  identified the  ``{\bf flavoring}'' of
Pomeron~\cite{Flavoring1}\cite{Flavoring4} as the primary dynamical mechanism for resolving the
paradox. A proper implementation of final-state screening correction, (or {\bf final-state
unitarization}), assures a unitarized ``gap distribution", with flavoring setting the relevant energy
scale.   We find that initial-state screening remains unimportant, consistent with the pole dominance
picture for elastic and total cross section hypothesis at Tevatron energies.   In fact, we shall
concentrate in the present discussion mostly on the flavoring and will comment only briefly on
screening. (A more complete treatment of final-state screening can be found in the Ref. 1 , and we
find that the effect turns out to be small.)

\section{Soft Pomeron at Low Energies}

In order to be able to answer the questions we have posed, it is necessary to first
provide a dynamical picture for a  soft Pomeron and to briefly review the notion of
``Harari-Freund" duality.

\subsection{Harari-Freund Duality}

Although Regge phenomenology pre-dated QCD, it is  important to recognize that it can be
understood as a phenomenological realization of non-perturbative QCD in a ``topological expansion",
e.g., the large-$N_c$ expansion. In particular, an important feature of a large-$N_c$ expectation is
emergence of the Harari-Freund two-component picture.~\cite{Harari_Freund}

For $P_{lab}\leq 20$ GeV/c, it was recognized that the imaginary part of any hadoronic two-body
amplitude can be expressed approximately as the sum of two terms:
$$ Im A(s,t)=R(s,t) + P(s,t).$$
From the s-channel point of view, $R(s,t)$ represents the contribution of s-channnel resonance while
$P(s,t)$ represents the non-resonance background. From the t-channel point of view, $R(s,t)$
represents the contribution of ``ordinary" t-channnel Regge exchanges and $P(s,t)$ represents the
diffractive part of the amplitude given by the Pomeron exchange.
Three immediate consequences of this picture are:
\begin{itemize}
\item{(a)}  {Imaginary parts of amplitudes which show no resonances should be dominated by Pomeron
exchange, ($R\simeq 0$, and $P\simeq constant$).}
\item{(b)} {Imaginary parts of $A(s,t)$ which have no Pomeron term should be dominated
by s-channel resonances,}
\item{(c)} {Imaginary parts of amplitudes which do not allow Pomeron exhange and show no resonances
should vanish,} 
\end{itemize}
Point (b) can  best be illustrated by  partial-wave projections of  $\pi
N\rightarrow
\pi N$ scattering amplitudes from  well-defined t-channel isospin exchanges. 
Point (c)  is best illustrated by examining the $K^+p\rightarrow K^0p$, where, by optical
theroiem,
$Im A(K^+p\rightarrow K^0 n)\propto \sigma_{tot}(K^+p)-\sigma_{tot} (K^0 n).$
The near-equality of these two cross sections, from the t-channel exchange view point, reflects the
interesting feature of exchange degeneracy for secondary Reggeons.
Finally, let us come to the point (a). From the behavior of $\sigma_{\pi^{\pm} p}$, $\sigma_{K^{\pm}
p}$,
$\sigma_{pp}$ and
$\sigma_{\bar p p}$, one finds that the near-constancy for the $P$-contribution corresponds to having
an effective ``low-energy" Pomeron intercept at 1, i.e.,  
$$\alpha_{\cal P}^{low}(0)\simeq 1.$$

\subsection{Shadow Picture and Inelastic Production}

 A complementary treatment of Pomeron at low energies is through the analysis of
inelastic production, which is responsible for the non-resonance background mentioned earlier.
Diffraction scattering as the shadow of inelastic production has been a well established mechanism
for the occurrence of a forward peak. Analyses of data up to ISR energies have revealed that the
essential feature of nondiffractive particle production  can be understood in terms of a
multipertipheral cluster-production mechanism. In such a picture, the forward amplitude at high
energies is predominantly absorptive and is dominated by the exchange of a ``bare Pomeron".  

In a
``shadow" scattering picture, the ``minimum biased" events are predominantly ``short-range ordered"
in rapidity and the production amplitudes can be described by  a multiperipheral cluster model. Under a
such an approximation to production amplitudes for  the right-hand side of an elastic unitary
equation,
$Im T(s,0)=
\sum_n |T_{2,n}|^2$, one finds that the resulting elastic amplitude is  dominated by the exchange of a Regge
pole, which we shall provisionally refer to as the ``bare Pomeron".
Next consider singly
diffractive events. We assume that the ``missing mass" component corresponds to  no gap events, thus the distribution is again
represented by a  ``bare Pomeron". However, for the gap distribution, one would insert the ``bare
Pomeron" just generated into a production amplitude, thus leading to the classical triple-Pomeron
formula.

Extension of this procedure leads to a ``perturbative" expansion for the total cross
section in the number  of bare Pomeron exchanges along a multiperipheral chain.  Such a framework was
proposed long time ago,~{\cite{FrazerTan}} with the understanding that the picture can make sense
at moderate energies, provided that the the intercept of the Pomeron is near one,
$\alpha_{\cal}(0)\simeq 1$, or less.

However, with the acceptance  of a Pomeron having an intercept
greater than unity, this expansion must be embellished or modified. It is quite likely that the
resolution for Dino's paradox lies in understanding how such an effect can be accomodated  within
this framework, consistent with the Pomeron pole dominance hypothesis.

\subsection{Bare Pomeron in Non-Perturbative QCD}

In a non-perturbative QCD setting, the Pomeron intercept is
directly related to the strength of the short-range order component of inelastic production and this
can best be understood in a large-$N$ expansion.~\cite{LeeVeneziano} In such a scheme,
particle production mostly involves emitting ``low-mass pions", and  the basic energy scale of
interactions is that of ordinary vector mesons, of the order of 
$1$ GeV.  In a one-dimensional multiperipheral realization for the ``planar
component" of the large-$N$ QCD expansion, the high energy behavior of a $n$-particle total cross
section is primarily controlled by its longitudinal phase space,
$\sigma_n\simeq (g^4N^2/{(n-2)!})(g^2N\log s)^{n-2} s^{J_{eff}-1}.$
Since there are only Reggeons at the planar diagram level, one has  $J_{eff}=2\alpha_R-1$ and, after
 summing over $n$,  one  arrives at Regge behavior for the planar component of $\sigma^{tot}$ where 
\be
\alpha_R=(2\alpha_R-1)+g^2N.
\label{eq:Planar}
\ee
At next level of cylinder topology,  the contribution to partial cross section increases due to its topological twists,
$\sigma_n\simeq {( g^4/ {(n-2)!})} 2^{n-2}(g^2N\log s)^{n-2} s^{J_{eff}-1},$
and, upon summing over $n$,  one arrives at a total cross section governed by a Pomeron exhange, 
$
\sigma_0^{tot}(Y)=g^4e^{\alpha_{\cal P} Y},
$
where the  Pomeron interecept is 
\be
\alpha_{\cal P}=(2\alpha_R-1)+2g^2N.
\label{eq:Cylinder}
\ee
Combining Eq. ({\ref{eq:Planar}) and Eq. (\ref{eq:Cylinder}), we arrive at an
amazing ``bootstrap" result,
$
\alpha_{\cal P}\simeq 1.
$

In a non-perturbative QCD setting, having a Pomeron intercept near 1 therefore depends crucially on
the topological structure of large-$N$ non-Abelian gauge
theories.~\cite{LeeVeneziano}
 In this picture, one has  $\alpha_R\simeq .5\sim .7$ and  $g^2N\simeq .3\sim .5  $.  With
$\alpha'\simeq 1$ $GeV^{-2}$, one can  also   directly relate 
$\alpha_R$ to the  average mass of typical vector mesons.
Since vector meson masses are controlled by constituent mass for light
quarks, and since constituent quark mass is a consequence of chiral symmetry breaking, the 
Pomeron and the Reggeon intercepts are directly related to fundamental issues in non-perturbative 
QCD. This picture is in accord with the Harari-Freund picture for low-energy Regge phenonemology.

Finally we note that, in a  Regge
expansion, the relative importance  of secondary trajectories  to the Pomeron is controlled by the
ratio
$e^{\alpha_{R}\> y}/e^{\alpha_{\cal P}\> y}= e^{-(\alpha_{\cal P}-\alpha_{\cal R})\> y}$. It follows 
that there exists a natural  scale in rapidity, $y_r$, $(\alpha_{\cal P}-\alpha_R)^{-1}<y_r \simeq
3\sim 5$.
The importance  of this scale $y_r$ is of course well known:  When using a
Regge expansion for total and two-boby cross sections, secondary trajactory contributions become
important and must be included whenever rapidity separations are  below $3\sim 5$
units. This scale  of course is also important for the triple-Regge region: There are two relevant
rapidity regions: one associated with the ``rapdity gap", $y\equiv \log \xi^{-1}$,  and the
other for the missing mass, $y_m\equiv \log M^2$.

\subsection{Conflict with Donnachie-Lanshoff Picture}
It has become increasingly popular to use the Dannachie-Landshoff picture~\cite{Donnachie_Landshoff}
where Pomeron intercept above one, i.e., $\epsilon\sim 0.1$. Indeed, it is impressive that various
cross sections can be fitted via Pomeron pole contribution over the entire currently available energy
range. However, it should be pointed out that Donnachie-Lanshoff picture is not consistent with the
Harari-Freund picture at low energies. 

It can be argued that the difference between these two approaches should not be important at high
energies. This is certainly correct for total cross sections. However, we would like to stress that
this is not the true for diffractive dissociation, even at Tevatron energies. This can best be
understood in terms  of rapidity variables,  $y$ and $y_m$.
Since $y+y_m\simeq Y$, $Y\equiv \log s$, it follows that,
even at Tevatron energies, the rapidity range for either $y$ or $y_m$ is more like that for a
total cross section at or below the ISR energies. Therefore, details of diffractive dissociation
cross section at Tevatron would depend on how a Pomeron is treated at releatively low subenergies.

\section{ Soft Pomeron and Flavoring}

  Consider for
the moment the following scenario where one has two different fits to hadronic total cross sections:
\begin{itemize}
\item
(a) ``High energy fit": $\sigma_{ab}(y)\simeq \beta_a\>\beta_b \>e^{\epsilon\> y}$ \hskip50pt for
\hskip50pt
$y>>y_f$,
\item
(b) ``Low energy fit": $\sigma_{ab}(y)\simeq \beta^{low}_a\>\beta_b^{low} $ \hskip50pt for
\hskip45pt $y<<y_f$.
\end{itemize}
That is, we envisage a situation where the ``effective Pomeron intercept", $\epsilon_{eff}$, 
increases from $0$ to $\epsilon\sim 0.1$ as one moves up in energies.   In order to have  a
smooth interpolation between these two fits, one can obtain the following order of magnitude estimate
$
\beta_p\simeq e^{-{\epsilon\> y_f\over 2}} \beta_p^{low}.
$
 Modern parametrization for Pomeron residues typically
leads to values of the order $(\beta_p)^2\simeq  14\sim 17 $ mb. However, before the advent of the
notion of a Pomeron with an intercept greater than 1, a typical parametrization would have a value
$(\beta^{low}_p)^2\simeq 35\sim 40$ mb, accounting for a near constant Pomeron contribution at low
energies. This leads to an estimate of $y_f\sim 8$, corresponding to $ \sqrt s \sim 50$ GeV.
This is precisely the energy scale where a rising total cross section first becomes noticeable.

The scenario just described has been referred to as ``flavoring", the notion that the
underlying effective degrees of freedom for Pomeron will increase as one moves to higher
energies,~\cite{Flavoring1}   and it has provided a dynamical basis
for understanding the value of Pomeron intercept in a non-perturbative QCD
setting.~\cite{Flavoring4} In this scheme, in order to extend a Regge phenomenology to low energies, 
both the Pomeron intercept and the Pomeron residues are {\bf scale-dependent}. We shall 
review  this mechanism shortly. However, we shall  first introduce a scale-dependent formalism
where  the entire flavoring effect can be absorbed into a flavoring factor,
$R(y)$, associated with each Pomeron propagator.

\subsection {Effective intercept  and Scale-Dependent  Treatment}

In order to be able to extend  a Pomeron repesentation below the rapidity scale $y\sim y_f$, we
propose   the following {\bf scale-dependent}  scheme where we introduce a flavoring factor for each
Pomeron propagator.  Since each
Pomeron exchange is always associated with  energy variable
$s$, (therefore a rapidity variable
$y\equiv
\log s$), we shall  parametrize the Pomeron trajectory function as 
\be
\alpha_{eff}(t; y)\simeq 1+\epsilon_{eff}(y) +\alpha' t,
\label{eq: EffectivePomeronTrajectory}
\ee
where $\epsilon_{eff}(y)$ has the properties
\begin{itemize}
\item 
{}  $\epsilon_{eff}\simeq \epsilon \simeq 0.1  $ \hskip100pt for \hskip40pt  $y>>y_f$,
\item 
{} $\epsilon_{eff}\simeq \epsilon_o\equiv \alpha^{low}_{\cal P}-1\simeq 0 $ \hskip55pt for
\hskip40pt 
$y<< y_f$.

\end{itemize}
For instance, exchanging such an effective Pomeron leads to a contribution to the elastic cross
section 
$
T_{ab}(s,t)\propto s^{1+\epsilon_{eff}(y) +\alpha' t}.
$
This representation  can now be extended down to the region $y\sim y_r$. 
We shall adopt a particularly convenient parametrization for $\epsilon_{eff}(y)$ in the next
Section when we discuss phenomenological concerns.

To complete the story, we need also to account for the scale dependence of Pomeron residues. What we
need is an ``interpolating" formula between the high energy and low energy sets. Once a choice for
$\epsilon_{eff}(y)$ has been made, it is easy to verify that a natural choice is simply
$\beta_a^{eff}(y)=\beta_ae^{[\epsilon-\epsilon_{eff}(y)]y_f}$. It follows that the total
contribution from a ``flavored" Pomeron corresponds to the following low-energy modification
 $$
 T_{a,b}^{a,b}(y,t)\rightarrow  R(y)\>
T_{a,b}^{cl}(y,t), $$
where $ T_{a,b}^{cl}(y,t)\equiv \beta_a\beta_be^{(1+\epsilon+\alpha'_{\cal P}t)\> y}$ is the
amplitude according to a ``high energy" description with a fixed Pomeron intercept, and 
$$R(y)\equiv
e^{-[\epsilon-\epsilon_{eff}(y)](y-y_f)},$$
 is a ``flavoring" factor. The effect of this modification can best be illustrated via
Figure~\ref{fig:total_xs}.

This flavoring factor
should  be consistently applied as part of each  ``Pomeron propagator". With  the  normalization 
 $R(\infty)=1$, we can therefore leave the residues alone, once they have been determined by a
``high energy" analysis.  For our single-particle gap  cross section, since there are
three Pomeron propagators, the renormalization factor is given by the following product: 
$ Z=
R^2(y)R(y_m).
$
It is instructive to plot in Figure~\ref{fig:flavoring}  this combination  as a function of either $\xi$ or
$M^2$ for various fixed values of total rapidity,  $Y$.

\subsection{Flavoring of Bare Pomeron}

We have proposed sometime ago  that ``baryon pair", together with other ``heavy flavor" production,
provides an additional energy scale, $s_f=e^{y_f}$,  for soft Pomeron
dynamics,  and this effect
can be  responsible for the perturbative increase of the Pomeron intercept to be greater than unity, 
$\alpha_{\cal P}(0)\sim
1+\epsilon,\>\>\epsilon>0$. One must bear this additional energy scale in mind in working with a soft
Pomeron.~\cite{Flavoring4}
That is, to fully justify using a Pomeron with an intercept $\alpha_{\cal P}(0)>1$, one must restrict
oneself to energies
$s>s_f$ where heavy flavor production is no longer suppressed. Conversely, to extrapolate Pomeron
exchange to low energies below $s_f$, a lowered ``effective trajectory" must be
used. This feature of course is
unimportant for total and elastic cross sections at Tevatron energies. However, it is important for
diffractive production since both $\xi^{-1}$ and $M^2$ will sweep right through this energy  scale at
Tevatron energies.

Flavoring becomes important whenever  there is a further inclusion of effective degrees of freedom
than that associated with light quarks. This can again be illustrated by a simple one-dimensional
multiperipheral model. In addition to what is already contained in the Lee-Veneziano
model, suppose that new particles can also be produced in a multiperipheral
chain. Concentrating on the cylinder level, the partial  cross sections will be  labelled by
two indices, 
\be
\sigma_{p,q}\simeq  (g^4/ p!q!) 2^{p+q}(g^2N\log s)^{p} (g^2_{f}N\log
s)^{q}s^{J_{eff}-1},
\label{eq:FlavoringTotalCrossSection}
\ee
where $q$ denotes the number of clusters of new particles produced. Upon summing over $p$ and
$q$, we obtain a ``renormalized" Pomeron trajectory
\be 
\alpha^{high}_{\cal P}=\alpha^{low}_{\cal P}+ \epsilon,
\label{eq:NewPomeron}
\ee
where $\alpha^{low}_{\cal P}\simeq 1$ and $\epsilon\simeq  2g^2_{f}N$. That is, in a non-perturbative
QCD setting, the effective intercept of Pomeron is a dynamical quantity,  reflecting the effective
degrees of freedom involved in near-forward particle
production.\cite{Flavoring4}

If  the new degree of freedom involves
particle production with high mass, the longitudinal phase space factor, instead of $({\log
s})^q$, must be modified. Consider the situation of producing one $N\bar N$ bound state together
with pions, {\it i.e.}, $p$ arbitrary and  $q=1$ in Eq. (\ref{eq:FlavoringTotalCrossSection}). 
Instead of
$(\log s)^{p+1}$,  each factor  should be replaced by
$(\log(s/m^2_{eff}))^{p+1}$, where $m_{eff}$ is an effective  mass for the $N\bar N$ cluster.
In terms of rapidity, the longitudinal phase space factor becomes
$(Y-\delta)^{p+1}$,  where
$\delta$ can be thought of as a one-dimensional ``excluded volume" effect. 
For heavy particle production, there will be an energy range over which keeping up to $q=1$
remains a valid approximation. Upon summing over $p$, one finds that the additional contribution
to the total cross section due to  the production  of one heavy-particle cluster
is~\cite{Flavoring1} 
$
\sigma^{tot}_{q=1}\sim \sigma_0^{total}(Y-\delta)(2g^2_fN)\log
(Y-\delta)\theta(Y-\delta),
$
where $\alpha_{\cal P}^{low}\simeq 1$. 
Note the effective longitudinal phase space ``threshold
factor",
$\theta(Y-\delta)$, and, initially, this term  represents a small perturbation  to the total
cross section obtained previously, (corresponding to
$q=0$ in Eq. (\ref{eq:FlavoringTotalCrossSection})), $\sigma_0^{total}$. 
Over a rapidity range, $[\delta, \delta+\delta_f]$, where $\delta_f$ is the average rapidity
required for producing another heavy-mass cluster, this is  the only term needed for incorporating
this new degree of freedom. As one moves to higher energies,
``longitudinal phase space suppression" becomes less important and more and more heavy particle
clusters will be produced. Upon summing over
$q$, we would obtain a new total cross section, described by  a  renormalized Pomeron,  with a new
intercept given by Eq. (\ref{eq:NewPomeron}).

We  assume that, at Tevatron, the energy is high enough so that this kind of ``threshold"
effects is no longer important.  How low an energy do we have to go before one encounter these
effects? Let us try to answer this question by starting out from low energies. As we have stated
earlier, for
$Y> 3\sim 5$, secondary trajectories become unimportant and using  a Pomeron with
$\alpha_{\cal}\simeq 1$ becomes a useful approximation. However, as new flavor production becomes
effective, the Pomeron trajectory will have to be renormalized. We can estimate for the relevant
rapidity range when this becomes important as follows:  $y_f >  2
\delta_{0}+
<q>_{min} \delta_f$. The first factor
$\delta_{0}$ is associated with leading particle effect, i.e., for proton, this is primariy due
to pion exchange.
$\delta_f$ is the minimum gap associated with one heavy-mass cluster  production, {\it  e.g.},
nucleon-antinuceon pair production. We estimate
$\delta_{0}\simeq  2$ and $
\delta_f\simeq 2\sim 3 $,  so that, with $<q>_{min}\simeq 2$,  we  expect the
relevant flavoring rapidity scale to be 
$y_f\simeq 8\sim 10$.

\section{ A Caricature of High Energy Diffractive Dissociation}

Both the  screening function and the flavoring function depend on the effective Pomeron intercept,
and  we shall adopt the following  simple parametrization.  The transition from
$\alpha^{low}(0)=1+\epsilon_o$ to
$\alpha^{high}(0)=1+\epsilon$ will occur over a  rapidity range,
$(y_f^{(1)}, y_f^{(2)})$. Let $ y_f\equiv \half(y_f^{(1)}+ y_f^{(2)})$ and $\lambda_f
^{-1}\equiv 
\half(y_f^{(2)}- y_f^{(1)})$. Similarly, we also define $\bar \epsilon\equiv
\half (\epsilon+\epsilon_o)$ and  $\Delta\equiv \half(\epsilon-\epsilon_o)$. A convenient
parametrization for $\epsilon_{eff}$ we shall adopt is 
$\epsilon_{eff}(y) =[\bar \epsilon +{\Delta} \tanh {{\lambda_f}
(y-\bar y_f)}].
$
The combination 
$[{\epsilon-\epsilon_{eff}(y)}]$ can be written as $(2{\bar\epsilon)\> [1+({s/
s_f})^{2\lambda_f}]^{-1}}$ where $ s_f=e^{ y_f}$. We arrive at a simple parametrization for our
flavoring function
\be
R(s)\equiv \bigl({s_f\over s}\bigr)^{(2\bar\epsilon)\> [1+({s\over \bar s_f})^{2\lambda_f}]^{-1}}.
\label{eq:FlavoringFactor2}
\ee
With
$\alpha_{\cal P}^{low}
\simeq 1$, we have $\epsilon_o\simeq 0$, $\bar \epsilon\simeq \Delta \simeq \epsilon/2$, 
and we expect that  $\lambda_f
\simeq 1$ and
$ y_f \simeq 8\sim 10$ are reasonable range for these
parameters.~\cite{GlobalFlavoring}

The most important new parameter we have introduced for understanding high energy diffractive
production is the flavoring scale, $s_f=e^{y_f}$. We have motivated by way of a simple model to
show that a reasonable range for this scale is $y_f\simeq 8\sim 10$. Quite independent of
our estimate, it is possible to treat our proposed resolution phenomenologically and determine this
flavoring scale from experimental data.

It should be clear that
one is  not attempting to carry out a full-blown phenomenological analysis here. To do that, one must
properly incorporate other triple-Regge contributions, {\it e.g.}, the
${\cal PPR}$-term for the low-$y_m$ region, the $\pi\pi{\cal P}$-term and/or the ${\cal RRP}$-term 
for the low-$y$ region, etc., particularly for $\sqrt s \leq \sqrt {s_f}\sim 100\> GeV$.  There
are also ``interference terms, e.g., ${\cal RPP}$, to take into account. What we  hope to achieve is
to provide a ``caricuture" of the interesting physics involved in diffractive production at collider
energies through our introduction of  the flavoring factors.~\cite{GlobalFlavoring}   

 Let us begin by first examining  what we should  expect. Concentrate on  the triple-Pomeron vertex
$g(0)$ measured at high energies. Let us for the moment assume that it has also been meassured
reliably at low energies, and let us denote it as
$g^{low}(0)$. Our flavoring analysis  indicates that these two couplings are related by
$g_{\cal PPP}(0) \simeq e^{-({3\epsilon y_f\over 2})}g_{\cal PPP}^{low}(0).
$
With $\epsilon\simeq 0.08\sim  0.1$ and $y_f\simeq 8\sim 10$,
using the value $g_{\cal PPP}^{low}(0)=0.364\pm 0.025\>\> mb^{\half}$, we expect a value of
$0.12\sim 0.18\> mb^{\half}$ for $g_{\cal PPP}(0)$. Denoting the overall multiplicative constant for
our renormalized triple-Pomeron formula by
$
K\equiv {\beta^2_a(0)g_{\cal PPP}(0)\beta_b(0)/ 16\pi}.
$
With $\beta^2_p\simeq 16\> mb$, we therefore expect $K$ to lie between the range $.15\sim .25\>\>
mb^2$.

We begin  testing our renormalized triple-Pomeron formula by first  determining the overall
multiplicative constant $K$ by normalizing the integrated $\sigma^{sd}$ to the measured CDF $\sqrt
s=1800\> GeV$ value. With $\epsilon=0.1$,
$\lambda_f=1$, this is done for a series of values for $y_f=7,\>8,\>9,\>10$. We obtain 
respective values for $K=.24,\> 0.21,\> 0.18,\> 0.15,$ consistent with our flavoring expectation.
As a further check on
the sensibility of these values for the flavoring scales, we find for the ratio $\rho\equiv
\sigma^{sd}(546)/\sigma^{sd}(1800)$ the values $0.63, \> 0.65,\> 0.68,\> 0.72$ respectively. This
should be compared with the CDF result of 0.834.

Having shown that our renormalized triple-Pomeron formula does lead to sensible predictions for
$\sigma^{sd}$ at Tevatron, we can improve the fit by enhancing the $PPR$-term as well as
$RRP$-terms which can become important. Instead of introducing a more involved phenomenological
analysis,  we simulate the desired low energy effect by having
$\epsilon_o\simeq -0.06\sim -0.08$. A remarkably good fit results with 
$\epsilon=0.08\sim 0.09$ and  $y_f=9$.~\cite{GlobalFlavoring} This is
shown in Figure~\ref{fig:diff_xs}. The ratio
$\rho$ ranges from $0.78\sim 0.90$, which is quite reasonable. The prediction for $\sigma^{sd}$ at
LHC is $12.6\sim 14.8\> md$.

Our fit leads
to a triple-Pomeron coupling in the range of 
\be
g_{\cal PPP}(0)\simeq .12\sim .18 \>\>{mb}^{\half}, 
\ee
exactly as expected. Interestingly, the triple-Pomeron coupling quoted in Ref. 3
($g(0)=0.69
\>mb^{1\over 2}$) is actually a factor of 2 larger than the corresponding low energy
value.~\cite{Tan1} Note that this difference  of a factor of 5 correlates almost precisely with the
flux renormalization factor 
$N(s)\simeq 5$ at Tevatron energies.

\section{ Final Remarks:}

In Ref. 1, a more elaborated treatment has been caried out where both the flavoring and the
final-state screening effects were considered. We have shown, given Pomeron as a pole, the total
Pomeron contribution to  a singly diffractive dissociation cross section can in principle be
expressed as $ {d\sigma \over dtd\xi}=[S_i(s,t)][D_{a,{\cal P}}(\xi,t)] [\sigma_{{\cal P} b}
(M^2,t)],
$ and 
$
D_{a,{\cal P}} (\xi, t)= S_f(\xi,t)F_{{\cal P}/a} (\xi, t).
$
\begin{itemize}
\item
 The first term, $S_i$, represents initial-state screening correction.  We have  demonstrated that,
with a Pomeron intercept greater than unity and with  a pole approximation for total and elastic
cross sections remaining valid, initial-state absorption cannot be large.  We therefore can justify 
setting 
$S_i\simeq 1$ at Tevatron energies.  

\item The first crucial step in our alternative resolution to the
Dino's paradox lies in properly treating the final-state screening, $S_f(\xi,t)$.
We  have explained in an expanding disk framework that the final-state screening  sets in at a
rapidity scale determined by the flavoring scale, $y_f$, which correlates well with the mysterious
scale, $\xi_2$, of Schlein.

\item
 We have stressed that the dynamics of a soft Pomeron in a
non-perturbative QCD scheme requires taking into account the effect of ``flavoring", the notion that
the effective degrees of freedom for Pomeron is suppressed at low energies. As a consequence, we
find that 
$ F_{{\cal P}/a} (\xi, t)=R^2(\xi^{-1}) F^{cl}_{{\cal P}/a} (\xi, t)$ and  $ \sigma_{{\cal
P}b}(M^2,t)=R(M^2)\sigma_{{\cal P}b}^{cl}(M^2,t)$ where $R$ is the ``flavoring" factor discussed in
this paper. 
\end{itemize}

It should be stressed  that our discussion depends crucially on the notion of  soft Pomeron
being a factorizable Regge pole. This notion  has always  been controversial.
Introduced more than thirty years ago, Pomeron was identified as the leading Regge trajectory with
quantum numbers of the vacuum with
$\alpha(0)\simeq 1$ in order to account for the near constancy of the low energy hadronic total cross
sections. However, as a Regge trajectory, it was unlike others which can be identified by the
particles they interpolate. With the advent of QCD, the situation has improved, at least
conceptually. Through
large-$N_c$ analyses and through other non-perturbative studies, it is natural to expect
Regge trajectories in QCD as manifestations of ``string-like" excitations for bound states and
resonances of quarks and gluons due to their long-range confining forces. Whereas ordinary meson 
trajectories can be thought of as ``open strings" interpolating $q\bar q$ bound states, 
Pomeron corresponds to  a ``closed string" configuration associated with glueballs. However, the
difficulty of identification, presummably due to strong mixing with  multi-quark states,
has not helped the situation in practice. In a simplified one-dimensional multiperipheral
realization of large-N QCD, the non-Abelian gauge nature nevertheless managed to re-emerge
through its topological structure.~\cite{LeeVeneziano}

 The
observation of ``pole dominance" at collider energies has hastened the need to examine more
closely various  assumptions made for Regge hypothesis from a more fundamental viewpoint. It is
our hope that by  examining Dino's paradox carefully and by  finding an alternative resolution to the
problem without deviating drastically from accepted guiding principles for hadron dynamics, Pomeron
can continued to  be understood as a Regge pole in a non-perturbative QCD setting. The resolution
for this paradox could therefore lead to a re-examination of other interesting questions from a
firmer theoretical basis.  For instance,  to be able to relate quantities such as the  Pomeron
intercept  to non-perturbative physics of color confinement represents a theoretical challenge of
great importance. 

{\bf Acknowledgments:}
I would like to thank  K. Goulianos for first getting me interested in this problem. I
am also grateful to  P. Schlein for explaining to me details of their work.
   Lastly, I  appreciate
the help from K. Orginos for both  numerical analysis and the preparation for
the figures. This work is supported in part  by the D.O.E.  Grant
\#DE-FG02-91ER400688, Task A.


\begin{figure}
$$
\epsfxsize=\textwidth
\epsfbox{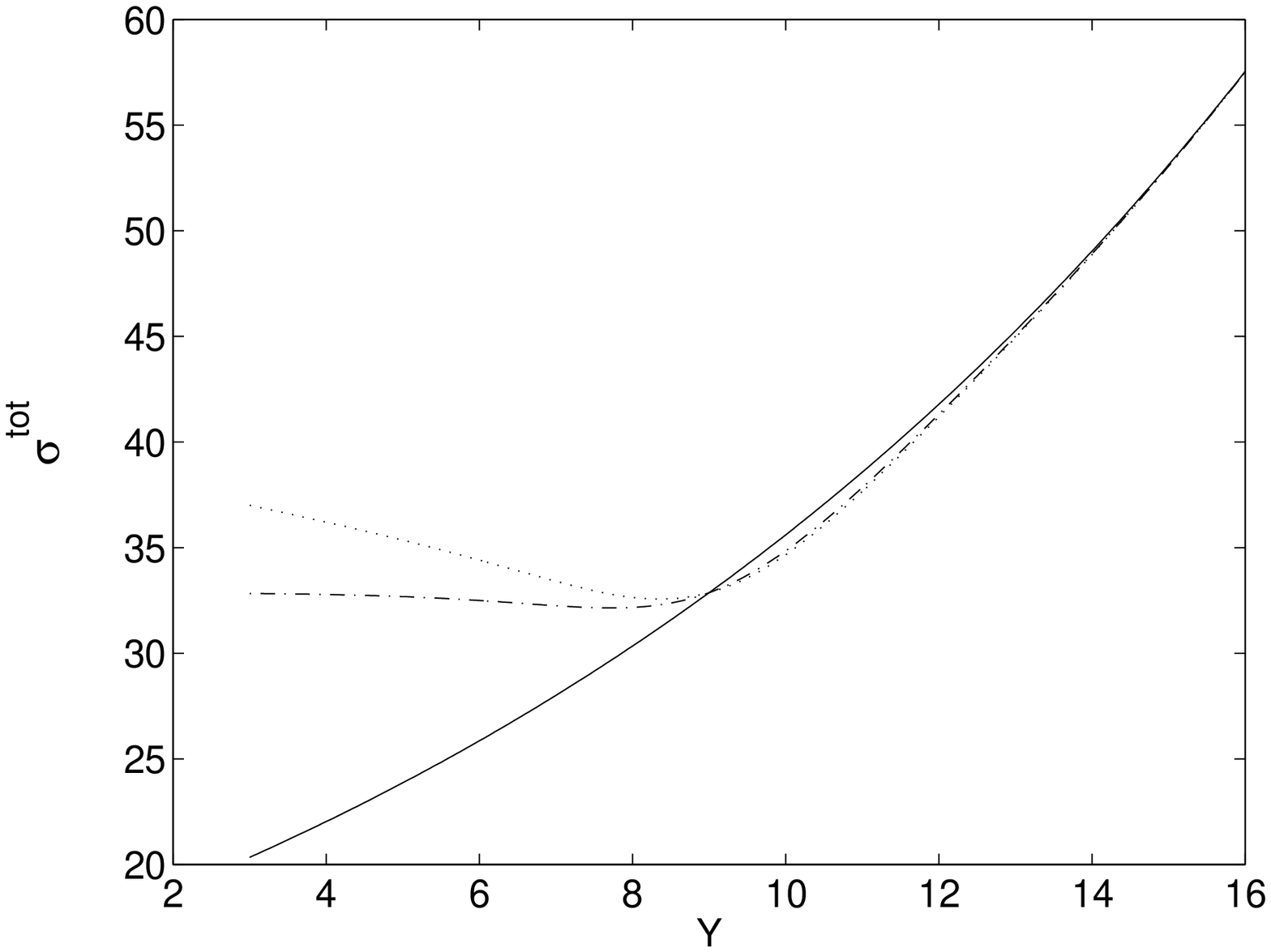} 
$$
\vspace{.1cm }
\caption{Effect of flavoring factor $R(s)$ when applied to a standard  rising cross
section: $\sigma^{cl}=\beta^2\> s^{\epsilon}$,  $\epsilon=0.1$ and  $\beta^2=16\> mb$, given by the
solid curve. With $R(y)$ given by  Eq.
(\ref{eq:FlavoringFactor2}), the 
 dashed-dotted curve has   $\epsilon_o=0$, 
$\lambda_f=1$, and flavoring scale $y_f=9$, and the dotted curve
corresponds to 
$\epsilon_o=-0.04$.} 
\label{fig:total_xs} 
\end{figure}

\begin{figure}
$$
\epsfxsize=\textwidth
\epsfbox{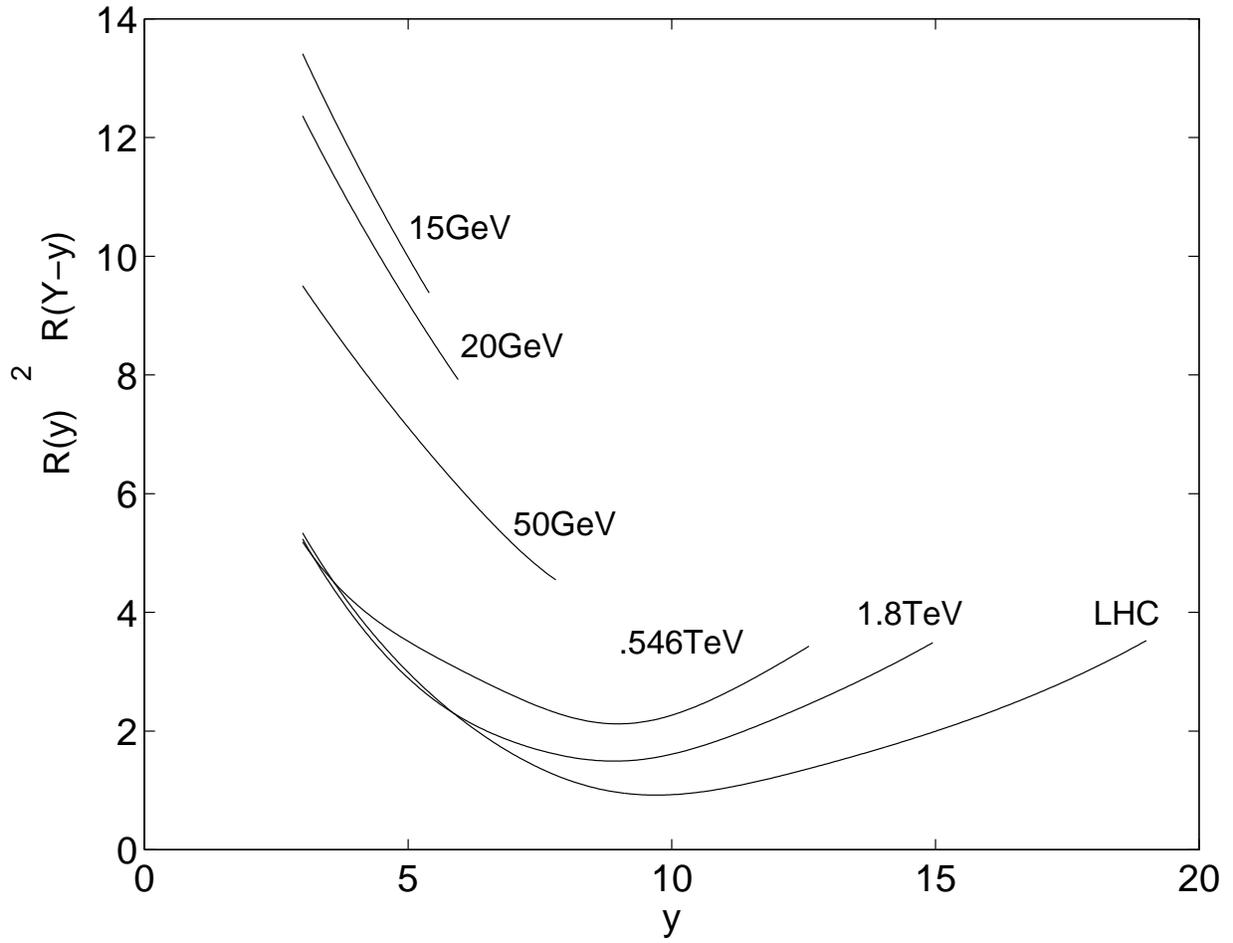} 
$$
\vspace{.1cm }
\caption{Renormalization factor due to flavoring alone, 
$Z_f(\xi;s)\equiv R^2(\xi^{-1})R(M^2)$,
as a function of rapidity $y=\log \xi^{-1}$ for various fixed center of 
mass energies.  These curves  correspond to parameters used for the solid
line in Figure 3.}  
\label{fig:flavoring} 
\end{figure}

\begin{figure}
$$
\epsfxsize=\textwidth
\epsfbox{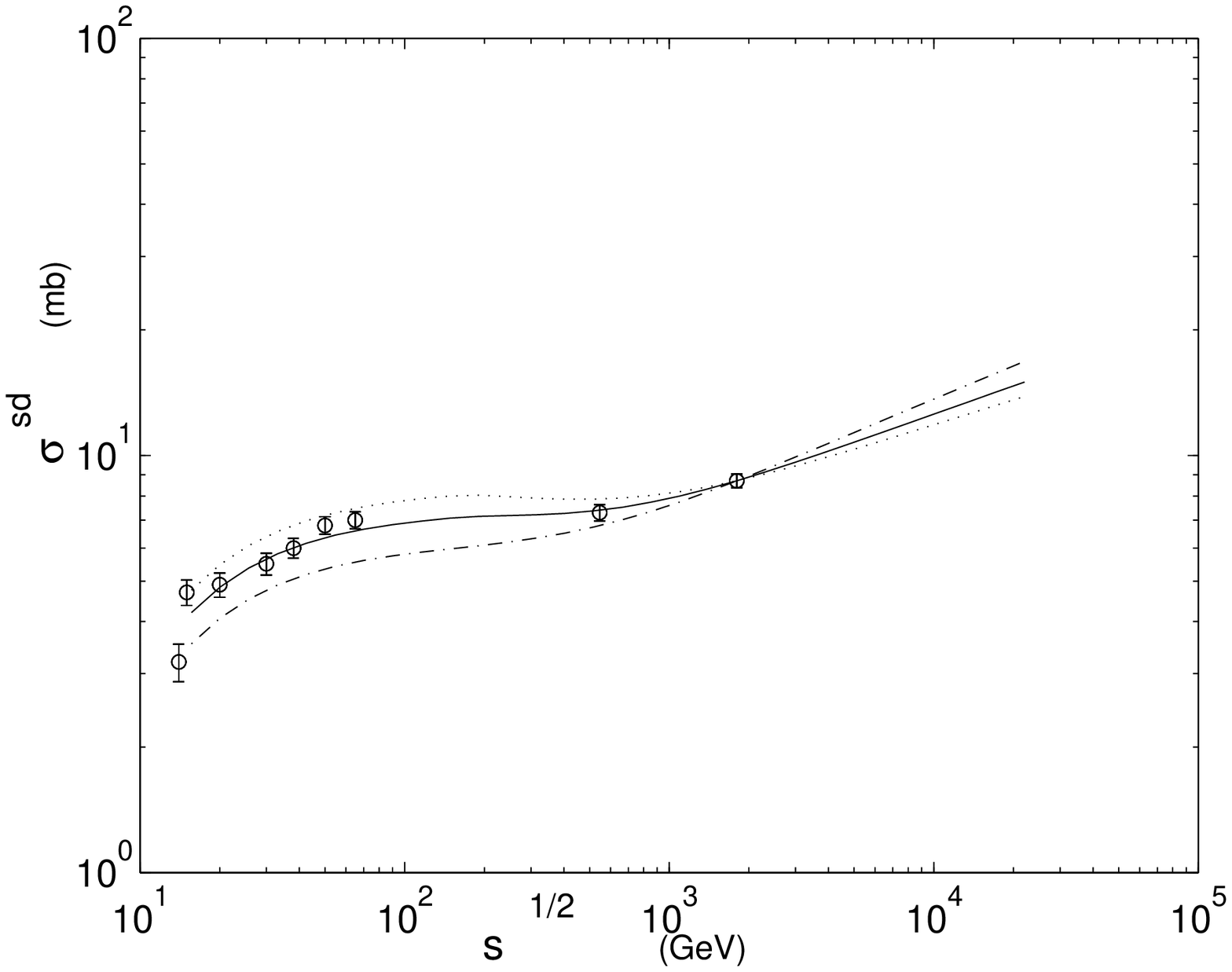} 
$$
\vspace{.1cm }
\caption{Various fits to representative single diffraction dissociation cross
sections extracted from Ref. 3 from ISR to Tevatron. The solid line and the 
dotted curve  correspond to   $\epsilon=0.08$, $\epsilon_o=-0.07$, 
$\lambda_f=1$,  $y_f=9$, with small amount of  final-state screening.  
 The dashed-dotted curve corresponds to no screening.}
\label{fig:diff_xs} 
\end{figure}


\begin{thebibliography}{99}

\bibitem{Tan1}  For a more detailed discussion, see: Chung-I Tan, ``{ Diffractive Production at
Collider Energies: Soft Diffraction and Dino's Paradox"}, 
hep-ph/9706276.


\bibitem{Classical}{We use
								$s_0= 1$ $ GeV^2$ as the basic energy scale through out this paper. These ``classical"
						expressions are: 
$F_{{\cal P}/a}^{cl} (\xi, t)=(1/16 \pi)\beta_a(t)^2 (\xi^{-1})^{2\alpha_{\cal P}(t) -1},$
 and
$\sigma_{{\cal P}b}^{cl}(M^2,t)=g(t) (M^2)^{\alpha_{\cal P}(0)-1}\beta_b(0).$  The triple-Pomeron
coupling $g(t)$ can be in principle determined by data below $\sqrt s\leq 30\> GeV$ where cross
sections are relatively insensitive to the choice of the Pomeron intercept value used.  }

\bibitem{Dino1} K.Goulianos, Phys. Lett., {\bf B} 358 (1995) 379.




\bibitem{Schlein1} P. Schlein, Proceedings of the 3rd Workshop on Small-x and
Diffractive Physics, Argonne National Laboratory, USA, September 1996.



\bibitem{GotsmanLevinMaor} E. Gotsman, E. M. Levin, and U. Maor, Phys. Rev., {\bf
D49} (1994) 4321.


\bibitem{Flavoring1} T. K. Gaisser and C-I Tan, Phys. Rev., {\bf D8} (1973) 3881;  C-I Tan, Proc. IX
Rencontres de Moriond, Meribel, France (1974).
We include both $N\bar N$ and $c\bar c$ production as well as
 other effects. The effective degrees of freedom involved are ``diquarks" and charm
quarks respectively. For color counting, a baryon is considered as a bound state of a quark and
diquark. In a more modern approach, baryons are to be considered as skerymions in a large $N$
appraoch. Again, they should be treated as independent degrees of freedom from mesons.

\bibitem{Flavoring4} C-I Tan, Proc. of 2nd International Conference on Elastic and Diffractive
Scattering, ed. K. Goulianos, (Editions Frantieres, 1987) p347;  C-I Tan,  Proc. of XIXth
International Symposium on Multiparticle Dynamics, Arles, ed. D Schiff and J. Tran T. V. (Editions
Frontieres, 1988) p361. We do not include ``semi-hard" production in the current
treatment for soft Pomeron. Flavoring will indeed be the primary mechanism in our construction of a
``Heterotic Pomeron". 

\bibitem{Harari_Freund} H. Harari, Phys. Rev. Lett. {\bf 20} (1968) 1395; P. G. O. Freund, Phys.
Rev. Lett. {\bf 20}, 235 (1968).

\bibitem{FrazerTan} W. Frazer, D. R. Snider and  C-I Tan, Phys. Rev.,  {\bf D8} (1973) 3180.

\bibitem{LeeVeneziano} H. Lee, Phys. Rev. Lett., {\bf 30} (1973) 719; G. Veneziano, Phys. Lett.,
{\bf B} 43 (1973) 314. See also, F. Low, Phys. Rev. {\bf D} 12 (1975) 163. A phenomnenological
realization of QCD emphasizing the topological structure of large-N gauge theories is Dual Parton
Model, (DPM). For a recent review, see: A. Capella, U. Sukhatme, C-I Tan, and J. Tran T. V.,
Physics Reports, {\bf 236} (1994) 225. 

\bibitem{Donnachie_Landshoff} A. Donnachie and P. V. Landshoff, Phys. Lett. {\bf B296} (1992) 227.
J. R. Cuddel, K. Kang, and S. K. Kim, Phys. Lett. {\bf B395} (1997) 311.

\bibitem{GlobalFlavoring} By choosing $\epsilon^{old}< 0
$, it is possible to provide a global ``average" description mimicking ``secondary trajectory"
contributions for various low energy regions. Acceptable estimates~\cite{Flavoring1} are
$\epsilon_o\simeq -0.11\sim -0.5$.


\end{thebibliography}
\end{document}